\documentclass[12pt]{JHEP3}

\def\R{{\mathbb R}}
\def\CA{{\mathcal A}}

\def\CO{{\mathcal O}}

\def\CO{{\mathcal O}}

\def\const{\hbox{const}}
\def\oh{\hbox{$\frac{1}{2}$}}

\def\ts{\otimes}
\title{Dynamical Noncommutativity}
\author{Andrzej Sitarz \thanks{Supported by Marie Curie Fellowship.} \\ 
Laboratoire de Physique Theorique, Universit\'e Paris-Sud,
Bat. 210, 91405  ORSAY Cedex, France \\
Institute of Physics, Jagiellonian University, Reymonta 4, 
30-059 Krak\'ow, Poland \\
E-mail:Andrzej.Sitarz@th.u-psud.fr}
\preprint{LPT-ORSAY 02-45}
\abstract{We present a model of Moyal-type noncommutativity with 
time-depending noncommutativity parameter and the exact gauge
invariant action for the $U(1)$ noncommutative gauge theory. 
We briefly result the results of the analysis of plane-wave 
propagation in a regime of a small but rapidly changing 
noncommutativity.}

\begin{document}

\section{Introduction}
There are many arguments that the picture of smooth space-time 
should change when approaching the Planck scale. Similarly as in the 
case of classical and quantum physics, when classical limit is recovered 
from the quantum description for macroscopic objects, we are probably 
in a need of a similar approach to describe in a consistent way 
the quantum theory of gravity. Apart from the model-independent 
considerations, such as black hole formation limits in quantum 
mechanics \cite{Fredenhagen} leading to uncertainty relations
for spacetime coordinates,  there are interesting models in which 
noncommutativity appears explicitly (see \cite{Wit} and reviews 
\cite{SN,Are} for an exhaustive list of string-related  references).

It should be stressed that there is more to noncommutative geometry 
than the particular type of the Moyal deformation, which appears in 
the above mentioned string-motivated theories. In particular, quite
appealing are the appearances of finite noncommutative geometry in
the Standard Model, where the Higgs field is recovered as the 
connection field in the finite-geometry \cite{Connes}. Matrix 
geometries like fuzzy manifolds are another possibility both 
as approximations or effective geometries at some energy 
scales \cite{GMS}. On the other hand, quantum deformations 
(see, for instance, \cite{Majid} for a review of links with physics) 
offer a vast realm of models with well-defined deformations of 
symmetries. 

In this paper we shall suggest a simple model, which extends 
the idea of a static or a global noncommutativity towards the 
time-dependent or a local one. Clearly, in general, such models 
will not provide an effective description of a flat space time. 
However, we believe that while considering real physical models, 
which are not static, like, for example, in cosmology, in black hole 
formation or even particle interactions, we need to consider a 
possibility that the effective noncommutativity, which might arise 
in the picture could be as well space- and time-dependent.

\section{Moyal deformation and its generalizations}

We shall begin by recalling different formulations 
of the Moyal deformations of $\R^4$ (though clearly it 
could be realised in any dimension).

On the level of generators the algebra could be described 
as defined by (selfadjoint, at least formally) 
generators $x^\mu$ and relations:
\begin{equation}
[x^\mu, x^\nu] = i \theta^{\mu\nu}, \label{def-1}
\end{equation}
where $\theta^{\mu\nu}$ is a real, constant, antisymmetric matrix. 
This provides an algebraic description, though restricts it, basically,
to the polynomial algebra. Therefore, it is more convenient to 
consider the deformation on the level of smooth ($C^\infty$) 
functions, where it becomes a nonlocal deformation of the usual 
product on the vector space of functions:
\begin{equation}
f(x) \ast g(x) = 
e^{\frac{1}{2} i \partial_\mu^x \theta^{\mu\nu} \partial_\nu^y} 
f(x) g(y)_{|_{x=y}}.
\label{def-2}
\end{equation}
Although for practical reasons one works mostly with smooth 
functions we may as well find out \cite{Rieffel} that the deformation 
is defined as well for the continuous (and bounded) functions, using 
oscillatory integrals:
\begin{equation}
f(x) \ast g(x) 
= \int_{\R^4} \int_{\R^4} d^4z \, d^4y 
\, f(x+\theta(z) ) g(x+y)  e^{2\pi i (y \cdot x)}. 
\label{def-3}
\end{equation}
where $\theta(z) = \theta^{ij} z_j$ and $x \cdot y$ is the 
standard scalar product of two vectors.

It is a nice exercise that both (\ref{def-2}) and (\ref{def-3}) 
give (\ref{def-1}) when applied to monomials.

The differential structures remain almost the same as in the 
undeformed case, in fact one can extend the linear isomorphism
between the deformed and undeformed functions onto the 
entire differential complex.

\subsection{The time-dependent noncommutativity}

It was observed \cite{Wit,Sch} that in the string theory the effective
geometry of space time generated by strings in the background 
of a constant $B$-field yields the Moyal deformation. As this
corresponds to a flat brane embedded in a flat background space 
one may ask what picture might arise from considering curved
branes or curved backgrounds.  The modifications of the Moyal
star product in this case were studied in details in \cite{CorSch},
where the it was shown that in the topological limit the deformation
is given by the Konstevich star product in the symplectic case 
or by the nonassociative version of the product in the most 
general situation:

\begin{equation}
\begin{array}{l}
f \ast g = f\, g + \frac{1}{2} \alpha^{ab} \partial_a f \, \partial_b g - \frac{1}{8}
\alpha^{ac} \alpha^{bd} (\partial_a \partial_b f) (\partial_c \partial_d g) \\
\phantom{xxxxx} - \frac{1}{12} \alpha^{ad} \partial_d \alpha^{bc} 
(\partial_a \partial_b f\, \partial_c g - \partial_b f\, \partial_a \partial_c g)
+ \CO(\alpha^3),
\end{array} \label{defK}
\end{equation}

where $\alpha^{ab}$ depends on the combination of $B$ and $F$ fields
on the brane \cite{CorSch}. Generally the deformation (\ref{defK})
is quite complicated, however, there exist some special cases in which
the modification of the commutation relations is minor. 

One of the simplest possible models arises when we assume that
$\alpha^{ab}$ depends only on one variable and, in addition, its 
component  in this direction vanishes:
$$ \alpha^{ab} \partial_b \alpha^{cd} = 0. $$

A particular solution of this condition is:
\begin{eqnarray*}
\alpha^{\mu 0} = \alpha^{0 \mu} = 0, \\ 
\partial_i \alpha^{jk} = 0,
\end{eqnarray*}
which leads to the construction of a time-dependent
noncommutativity of space coordinates:
\begin{equation}
\begin{array}{l}
{}[x^i, x^j] = i \theta^{ij}(t), \\
{}[x^i, t] = 0,
\end{array} \label{nonc-t1}
\end{equation}
where $\theta^{ij}(t)$ is a smooth funcion of $t$ 
valued in antisymmetric real matrices. For arbitrary 
smooth functions (identified as elements of the 
vector space of the deformed algebra) the product 
could be written as: 
\begin{equation}
f(x) \ast g(x) = e^{\frac{1}{2} i \partial_i^x \theta^{ij}(t) 
\partial_j^y} f(x) g(y)_{|_{x=y}}.
\label{def-4}
\end{equation}

It could be argued that such "dynamical noncommutative manifolds" 
are as good object as the normal manifolds or deformed ones 
\cite{Sit}. They might give an insight as to whether noncommutavity 
could be treated within nontrivial dynamical systems and whether 
it might have evolved with the cosmological evolution or, for instance, 
it could accompany the creation of black holes. 

Let us consider functions on $\R^4$, which, satisfy (for the polynomials)
have the commutation relations (\ref{nonc-t1}). In our case, when
we consider a general construction, the function $\theta$ can 
{\em a priori} be arbitrary, in particular, it could interpolate 
between the highly noncommutative and commutative regime. 

What changes in comparison with the $\theta=\const$ case 
is the differential calculus:
\begin{equation}
\begin{array}{l}
{}t\, dx^i = dx^i \, t \\
{}x^i \, dt = dt\, x^i \\ 
{}[x^i, dx^j ] = \left( \oh i \dot{\theta}^{ij} + A^{ij}(t) \right) \, dt ,
\end{array} \label{nonc-f}
\end{equation}
where $A^{ij} = A^{ji}$ is any symmetric matrix. The change is 
only on the level of commutations between the forms and 
the functions, as the products of the generating forms 
do not change at all:
\begin{equation}
dx^\mu \, dx^\nu + dx^\nu \, dx^\mu = 0.
\end{equation}

The commutation rules between differentials and the arbitrary 
functions are:
\begin{equation}
f(x) \, dx^j = dx^j \, f(x) + \left( \oh i \dot{\theta}^{ij}+ A^{ij}(t) \right)\, (\partial_i f(x)) \, dt
\label{f-dx}
\end{equation}
We use the usual $\R^4$ partial derivatives $\partial_\mu$, however, note that 
due to noncomutativity (\ref{def-4}) they no longer must obey the Leibniz rule, 
in particular we observe that:
\begin{equation}
\partial_t (f \ast g) = (\partial_t f) \ast g + f \ast (\partial_t g) 
+ \left( \oh i \dot{\theta}^{ij}+ A^{ij}(t) \right) (\partial_i f) \ast (\partial_j g). 
\label{noLeib}
\end{equation}

As a final remark of this section let us observe that the standard
conjugation on the vector space of complex-valued smooth functions
on $\R^4$ is still a well defined conjugation of the deformed 
algebra: i.e:
\begin{equation}
\overline{(f \ast g)} = \bar{g} \ast \bar{f},
\end{equation} 
and that it extends on the differential algebra so 
that all $dx^\mu$ are selfadjoint.

It is easy to see that the considered time-dependent 
deformation have the same trace (integral) as the usual 
one, therefore we shall have:
\begin{equation}
\int f \ast g = \int d^4x \,f(x)g(x). 
\end{equation}
and 
\begin{equation}
\int (\partial_\mu f)  = 0. 
\end{equation}

Note that in the commutation relations for differentials there are two
terms of different origin: the derivative of $\theta^{ij}$, which is 
connected to the dynamical deformation of space-time and the term
proportional to an arbitrary symmetric matrix $A^{ij}(t)$. The latter
is in no way related to our deformation, hence we shall put 
$A^{ij}(t) = 0$ throughout the rest of this paper.

\section{Klein-Gordon equation}

Although there is no natural Hodge star on the differential algebra,
which we have constructed in the previous section  (at least not as 
a bimodule map, still a left-module Hodge map exists and is 
a straightforward generalisation from the undeformed case) we 
have a natural metric understood as a bimodule map 
$$g: \Omega^1(\CA) \ts_\CA \Omega^1(\CA) \to \CA, $$
given by (for instance):
$$ g(dx^i, dx^j) = - \delta^{ij}, \;\;\; 
g(dx^i, dt) = 0 = g(dt, dx^i), \;\;\; g(dt,dt)=1. $$

Then we might consider the dynamical scalar field action of the 
form 
\begin{equation}
S_\Phi = \int g(d\Phi, d\Phi^*) = 
\int \eta^{\mu\nu} (\partial_\mu \Phi) (\partial_\nu \Phi)^*,
\end{equation}

where $\eta$ denotes the tensor components of the above 
defined flat  Minkowski metric $(+,-,-,-)$. The resulting wave 
equation is formally the same:
\begin{equation}
\eta^{\mu\nu} \partial_\mu \partial_\nu \Phi = 0. 
\end{equation}
though we must take into account that $\partial_t$ is
not a derivation.

It appear, however that on functions, which depend only
on linear functions of spatial coordinates, that is:
$ f = f(t, k_i x_i) $ all partial derivatives are 
derivations, let us verify it on $(k_i x_i)^2$:
$$ \partial_0 (k_i x_i)^2 = \frac{1}{2} \dot{\theta}^{ij} k_i k_j = 0,$$
because $\theta$ is antisymmetric. Extending the results 
on all polynomials (by induction) we might conclude that
the canonical solution of the Klein-Gordon equation on the 
dynamical noncommututative deformation we present is the 
same as in the undeformed case $ \Phi = \Phi(k_\mu x^\mu)$
for any null vector $k_\mu$, $k^2=0$.

\section{Gauge theory}

The principal change in noncommutative gauge theory is the appearance 
of gauge-field self-interaction terms (nonlocal from the commutative 
point of view) for the $U(1)$ gauge field theory.  A noncommutative 
gauge connection is an antiselfadjoint one-form $A_\mu dx^\mu$, 
for which the gauge strength field is $F = dA + A\, A$. 

First of all,  unlike in the commutative case the components 
of the gauge potential shall not be imaginary functions: that shall 
be true only for the spatial components, whereas the time component
shall be composed of an arbitrary imaginary field and have
a part depending of the other fields:
\begin{equation}
A_0 + (A_0)^* = \frac{1}{2} i \dot{\theta}^{ij} (\partial_i A_j). 
\label{asa}
\end{equation}

Moreover, in the case of time-dependent noncommutativity we have, 
in addition to the appearance of the usual nonlinear terms, the problem 
of the proper definition of the gauge invariant Yang-Mills action.

Since, due to (\ref{f-dx}) functions do not commute with differentials,
the rules of gauge transformations for the components of the 
field strength are:
\begin{eqnarray}
&& F_{ij}' = U^\dagger \ast F_{ij} \ast U \\
&& F_{0i}' = U^\dagger \ast F_{0i} \ast U 
+ \oh i \dot{\theta}^{kj} U^\dagger \ast F_{ij} \ast (\partial_k U),
\end{eqnarray}

where
\begin{equation}
\begin{array}{l}
F_{ij} = \partial_i A_j - \partial_j A_i + [A_i, A_j], \;\;\; i<j,  \\
F_{0i} = \partial_0 A_i - \partial_i A_0 +
 [A_0, A_i]  - \oh i  \left( \dot{\theta}^{mn} A_n * (\partial_m A_i) \right). 
\end{array}
\end{equation}

Clearly, $F_{ij} F^{ij}$ remains gauge covariant but $F_{0i} F^{0i}$ 
is not.  However, let us observe that we can easily find a gauge 
covariant expression $\widetilde{F}_{0i}$:
\begin{equation}
F_{0i} - \oh i \dot{\theta}^{kj} F_{ij} \ast A_k.
\end{equation}

The additional term transforms like:
$$  \oh i \dot{\theta}^{kj} \ast U^\dagger \ast F_{ij} \ast A_k \ast U
+\oh  i \dot{\theta}^{kj} \ast U^\dagger \ast F_{ij} \ast (\partial_k U), $$
and exactly cancels the corresponding additional term in the 
transformation rule of  $F_{i0}$. Then the term, which would
contribute to the gauge-invariant action, $\widetilde{F}_{0i}$, 
reads:
$$ \widetilde{F}_{0i} = 
\partial_0 A_i - \partial_i A_0 + [A_0, A_i] 
- \oh i \dot{\theta}^{kj} \left( F_{ij} \ast A_k + 
A_j \ast (\partial_k A_i) \right).$$

Finally, we can propose the gauge invariant action:
\begin{equation}
S = \int \sum_{i<j} (F_{ij} \ast F_{ij}^*) - 
\sum_i (\widetilde{F}_{0i} \ast \widetilde{F}_{0i}^*),
\end{equation}

which, in addition to the standard terms of classical electrodynamics
contains the corrections resulting from the dynamical character of 
noncommutativity of space. We shall analyse these corrections in
more detail for the example of plane-waves.

\section{Physical effects in dynamical noncommutativity}

In this section we shall briefly discuss the potential observable
physical consequences of the dynamical noncommutativity. In what 
follows we shall always assume that the parameter $\theta$, which 
determines the strength of noncommutativity is negligible,  however, 
its time variation is not. Therefore we shall skip all the terms 
proportional to $\theta$ and keep only the lowest order corrections in 
$\dot{\theta}$.  This corresponds to the intuitive picture of short 
rapid "bursts" of noncommutative in the past history of the universe.
We shall concentrate on the perturbation to the propagation 
of electromagnetic waves caused by such events, using the
earlier derived action (4.6) and assuming that the gauge potential is 
of the plane-wave form to investigate the model.

First, note that we have to take into account the nonlinear form 
of the anti-selfadjoint condition (\ref{asa}), thus leading to:
\begin{equation}
\begin{array}{l}
 A_i = i p_i f( \omega t + k_i x^i ),\\
 A_0 = i p_0 f( \omega t + k_i x^i ) - \frac{1}{4} 
 \dot{\theta}^{ij} (k_i p_j) f'( \omega t + k_i x^i ).
 \end{array}
 \end{equation}
 where $p_\mu, k_i$ are fixed (real) vectors 
 and $\omega$ is a (real) constant.

For the components of the field strength we have:
\begin{equation}
\begin{array}{l}
F_{0i} = i (\omega p_i - k_i p_0) f' + \oh i f \ast f' \dot{\theta}^{jk} (k_j p_k p_i) 
+ \frac{1}{4}  f'' \dot{\theta}^{jk} (k_j p_k k_i), \\
F_{ij} = i (k_i p_j - p_i k_j) f'
\end{array}
\end{equation}

The gauge-covariant term obtained as a correction of $F_{i0}$ will be:
\begin{equation}
\begin{array}{l}
\tilde{F}_{0i} = F_{0i} 
- \frac{1}{2} i \dot{\theta}^{jk}  k_j p_i p_k f' \ast f \\
  \phantom{xxx} = i (\omega p_i - k_i p_0) f' - \frac{1}{2} i \dot{\theta}^{jk} 
  k_j p_i p_k  ( f' \ast f + f \ast f') \\
  \phantom{xxx} + \frac{1}{4}  f'' \dot{\theta}^{jk} (k_j p_k k_i).
 \end{array}
 \end{equation}
 
The action for the plane-wave Ansatz reads:
\begin{equation}
\begin{array}{l}
- \left( \vec{k}^2 \vec{p}^2 + 2(\omega p_0) (\vec{k} \vec{p})
- (\vec{k} \vec{p})^2 - \omega^2 \vec{p}^2 - p_0^2 \vec{k}^2 \right) \int (f')^2 \\
\phantom{xxx} -2  k_j p_k 
\left( \omega \vec{p}^2 - p_0 (\vec{k} \vec{p}) \right)  
\int \dot{\theta}^{jk}  f'*f'*f  \\
\phantom{xxx} + o(\dot{\theta}),
\end{array}
\end{equation}
where we have omitted terms of higher order in $\dot{\theta}$, and we have
used the trace property of the integral (2.12).

Let us see for a while what might be the consequences of the additional 
term, which appears in the effective action for the electromagnetic plane 
wave. First of all, the lowest order local term for the self-interaction is of
third order in the field $A$. It vanishes only if the wave is $\dot{\theta}$ 
polarised, that is if $\dot{\theta}^{ij} k_i p_j = 0$. Since the correction 
parameter is itself a function of time, of which we have assumed that its 
derivative is (relatively) significant we cannot treat the additional term 
as a constant modification,  which might modify the dispersion relation.
Instead the highly nonlinear character of the term suggests rather that
one would expect the dynamical noncommutativity to induce effects similar 
to these of nonlinear optics, and, in particular, the frequency doubling
of electromagnetic radiation, which is, in principle, measurable.

Above, we have taken only the simple example based on the pure 
$U(1)$ gauge theory in our model of dynamical noncommutativity, 
which we interpret as electrodynamic - to see whether there could be
some observable consequences.  As we can see, even on the pure
classical level the answer is positive: dynamical noncommutativity
adds some nontrivial corrections leading to nonlinear selfinteractions 
of electromagnetic potential, while still keeping the gauge invariance 
of the theory intact. Still, potentially even more interesting effects might 
appear on the level of interactions between light and matter, in 
particular, in the possible corrections to the atomic spectra coming 
from dynamical noncommutativity.

\end{document}